# Absence of spin-orbit-coupling-induced effects on lattice dynamics in CePt$_3$Si

S. Krannich[1], D. Lamago[1], D. Manske[2], E. Bauer[3], A. Prokofiev[3], R. Heid[1], K.-P. Bohnen[1] and F. Weber[1]

[1] Institut für Festkörperphysik, Karlsruher Institut für Technologie, P.O. 3640, D-76021 Karlsruhe, Germany
[2] Max-Plank-Institut für Festkörperforschung, Heisenbergstrasse 1, D-70569 Stuttgart, Germany
[3] Institute of solid state physics, Vienna University of Technology, A-1040 Vienna, Austria

**Motivated by model calculations for the heavy fermion superconductor CePt$_3$Si predicting phonon anomalies because of anti-symmetric spin-orbit coupling we performed a detailed experimental study of the lattice dynamical properties of CePt$_3$Si. In particular, we investigated the dispersion of transverse acoustic and low energy optic phonon branches along the [110] direction using inelastic neutron scattering. In these branches, we found deviations from our ab-initio lattice dynamical calculations, which overall give a good description of the phonon dispersion in CePt$_3$Si. However, the agreement for the [110] transverse modes can be improved if we neglect the Ce *4f* states, done in an additional calculation. We conclude that the lattice dynamics of CePt$_3$Si are conventional and that the observed deviations are not related to effects of anti-symmetric spin-orbit-coupling. More likely, ab-initio calculations overestimate the exchange between different phonon branches, particularly in the presence of *4f* electron states. Our results imply that the ASOC plays less a role in non-centrosymmetric superconductors than commonly believed.**

## I. Introduction

Even more than a hundred years after its discovery, superconductivity still is one of the most intense research fields in solid state physics demonstrated, e.g., by the tremendous amount of work on the newest unconventional superconductors, i.e. the iron-based ones. One feature that puts them apart from the well-known conventional superconductors is the proximity of the superconducting phase to magnetic order. Here, the heavy-fermion compound CeCu$_2$Si$_2$ discovered in 1979 was the first example for a close proximity of superconductivity and magnetism[1].

A new twist in the story of superconductivity appeared with the first non-centrosymmetric (NCS) heavy-fermion superconductor CePt$_3$Si, in which superconductivity ($T_c = 0.75$ K) coexists with antiferromagnetism ($T_N = 2.2$ K).[2] The new characteristic feature is the absence of an inversion centre, which results in a strong Rashba-type antisymmetric spin orbit coupling (ASOC) enabling the mixing of singlet and triplet pairing. Hence, CePt$_3$Si cannot readily be classified as singlet or triplet superconductor.[3] Indeed experimental observations point towards an order parameter of mixed symmetry.[2,4] Currently, we know a couple of NCS heavy-fermion superconductors,[5] CePt$_3$Si however remains the only one with a finite $T_c$ at ambient pressure.

Shortly after the discovery of CePt$_3$Si theoretical work on NCS superconductors has started focusing on symmetry considerations of the superconducting order parameter and phenomenological models [6]. A lot of work has been done on the static spin susceptibility [7] as well as of a microscopic paring theory based on such calculations [8]. Another important issue is impurity scattering and how it affects the superconducting state [9]. Furthermore, Andreev reflection and surface bound states have been addressed in Refs. [10]. Of particular theoretical interest in CePt$_3$Si and in Li$_2$Pt$_3$B are unusual vortex effects [11] as well as topological protected states [12]. Very recently, also the Josephson effect [13] as well as gauge modes and the Leggett mode have been investigated for various NCS compounds [14].

Recently, Klam *et al.* put forward a kinetic theory for NCS superconductors including effects of ASOC [15,16]. Employing a microscopic Hamiltonian to CePt$_3$Si they have calculated the underlying superconducting pairing interaction using the dynamical spin- and charge susceptibilities. Then, the singlet-to-triplet ratio is determined by the strength of the ASOC and the nesting properties of the band structure. Combining their results with *ab-initio* calculations of the lattice dynamical properties yielded a peaked static susceptibility for CePt$_3$Si, which should give rise to phonon anomalies. In general, these phonon anomalies are caused by electronic correlations and thus should reflect the pairing mechanism in the presence of spin-orbit coupling. Hence, a corresponding experimental investigation of the lattice dynamical properties could detect this particular effect of ASOC in CePt$_3$Si.

## II. Experiment

The neutron scattering experiments were performed on the 1T triple-axis spectrometer at the ORPHEE reactor at LLB, Saclay. Double focusing pyrolytic graphite monochromators and analyzers were employed for phonons with energies up to 20 meV. Higher energy phonons were measured with a Cu220 monochromator to achieve high resolution. A fixed analyzer energy of 14.7 meV allowed us to use a graphite filter in the scattered beam to suppress higher orders. The wave vectors are given in reciprocal lattice units (r.l.u.) of $(2\pi/a, 2\pi/b, 2\pi/c)$, where $a = b = 4.072$ Å and $c = 5.442$ Å. Measurements were carried out both in the [100]-[001] and in the [110]-[001] scattering planes of the simple tetragonal reciprocal lattice. Two high quality single crystals were used for the experiments weighing 1.5 g and 3 g, respectively. Both were grown by the Czochralski

## III. Results

Phonon measurements in CePt$_3$Si are complicated by the low symmetry of the crystal structure and the corresponding ambiguous phonon selection rules. Typical sets of raw data as shown in Figs. 1(a),(b) comprise a couple of excitations having energies in close vicinity to each other. The polarization of the observed phonon modes can be inferred qualitatively by the measured wave vector position. The data in Fig. 1 show c-axis polarized modes measured at $Q = (0.35, 0.35, 3)$ [Fig. 1(a)] and longitudinal modes at $Q = (1.45, 1.45, 0)$ [Fig. 1(b)] dispersing along the [110] direction.

The measured energies are summarized in panels *(c) – (e)* of Figure 1. Results obtained in the Brillouin zones (BZ) adjacent to $\tau = (0,0,3)$, $(1,1,0)$ and $(2,2,0)$ are shown in panel *(c)*. We note that these longitudinal and c-axis polarized modes are symmetrically not independent and therefore shown in one plot. Panel *(d)* features the dispersion of *a-b* plane polarized transverse modes, while energies of [100] longitudinal modes measured at $Q = \tau + (h, 0, 0)$, $\tau = (1,0,0)$ and $(2,0,0)$, are displayed in *(e)*.

In order to make a more detailed analysis, we performed lattice dynamical calculations using density-functional-perturbation theory[1] (DFPT, see details in the Appendix) and compared the observed spectral weight distributions with calculated structure factors for the respective scattering wave vectors **Q**. We find that the predicted number of phonon peaks, e.g. at $Q = (0.35, 0.35, 3)$ [Fig. 1(a)] and $(1.45, 1.45, 0)$ [Fig. 1(b)] is in good agreement with experiment. However, some calculated phonon energies, e.g. that for the lowest energy mode at $Q = (0.35, 0.35, 3)$ [Fig. 1(a)], are clearly off. Nevertheless, a clear assignment to particular phonon modes and symmetries was possible and we found generally good agreement for the investigated wave vectors . [Figs. 1(c)-(e)]. Our results demonstrate that DFPT is able to predict energies for phonons up to 20 meV in CePt$_3$Si. In contrast, the energies of two additional flat dispersion bands at calculated energies of 43 meV and 57 meV were found to have up to 15% lower energies. This difference might be explained by the 4% smaller unit cell volume in the calculation compared to experiment (see tab. 1). High energy optical modes have typically poalization patterns with strong opposite movements of the atoms and are therefore stronger affected by differences in the volume than low energy modes. As these high energy modes have mainly Si character and no particular impact of ASOC can be expected, we restrict the following discussion to the low energy dispersion.

Theory predicts phonon anomalies because of ASOC in low energy acoustic phonon modes [16]. The strongest effect was calculated for the *a-b*-plane posarized transverse acoustic (TA) mode along the [110] direction. Accordingly, we focused on the measurements in this direction [see Figs. 1(c),(d)]. Despite the overall good agreement of method[2]. The samples were mounted in a closed-cycle refrigerator and all measurements presented in the following were done at $T = 10$ K.

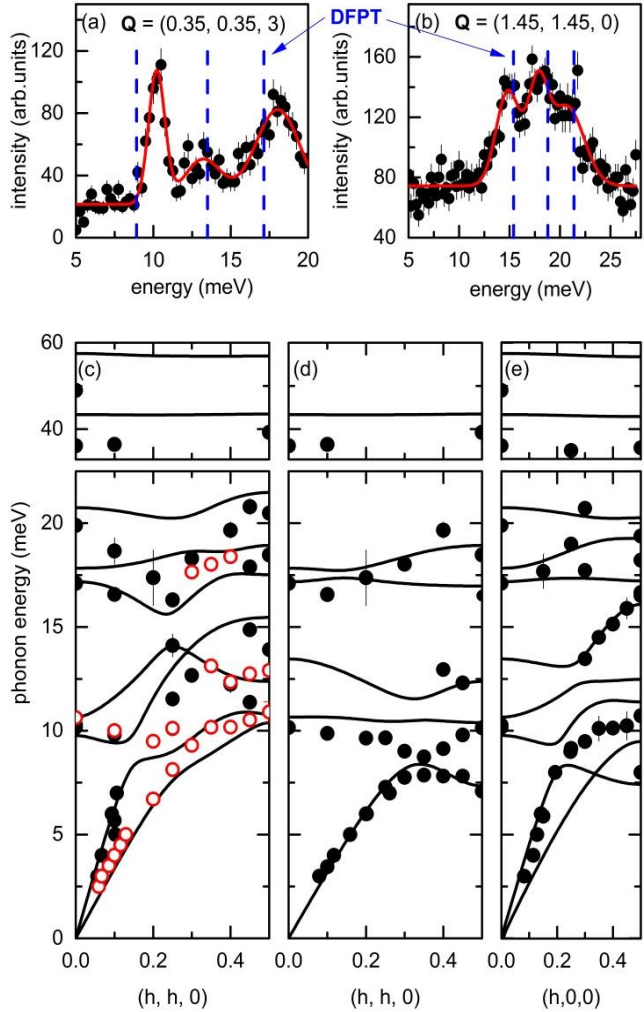

**Figure 1:** (color online) *(a)(b)* Typical sets of inelastic neutron scattering data at two different wave vectors and T = 10 K. The red line is a fit consisting of Gaussian peaks on a linear experimental background. The vertical dashed (blue) lines indicate the calculated energies of phonons with measurable intensities at the respective wave vectors. *(c)-(e)* Comparison of calculated (lines) and observed (dots) phonon dispersion in CePt$_3$Si for selected symmetries: *(c)* Longitudinal and c-axis polarized phonon branches dispersing along the [110] direction. Longitudinal modes were measured adjacent to $\tau =(1,1,0)$ and $(2,2,0)$ reciprocal lattice vectors (filled symbols). *c*-axis polarized modes were measured in the Brillouin zone adjacent to $\tau = (0,0,3)$ (open symbols). *(d) a-b*-plane transverse polarized modes along the [110] direction measured mostly in the Brillouin zones adjacent to $\tau = (2,2,0)$ and $(3,1,0)$. *(e)* Longitudinal phonon branches along the [100] direction were measured in the Brillouin zones adjacent to $\tau = (1,0,0)$ and $(2,0,0)$. Experimental error bars are, except for a couple of phonon energies, smaller than the symbol size. There are no phonon modes at 22.5 meV $\leq E \leq$ 33 meV.

---

[1] If not stated otherwise, calculations were performed without spin-orbit coupling.

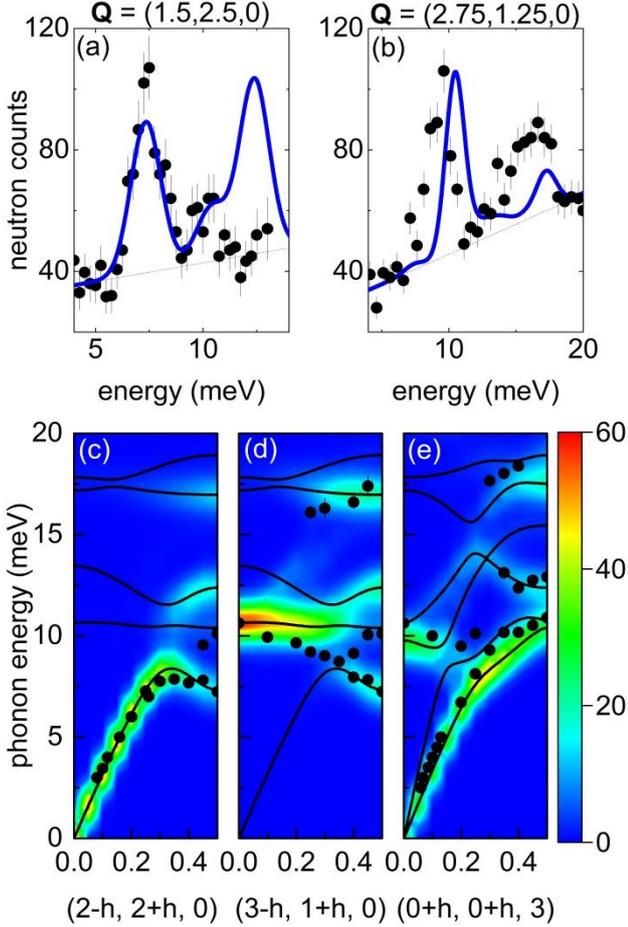

**Figure 2:** (color online) *(a),(b)* Comparison of raw data (dots) with calculated intensities (solid lines) for two wave vectors. The estimated experimental background (dashed line) was added to the calculation for an easier comparison. *(c)-(e)* Comparison of calculated phonon intensities (color code) and energy positions of observed peaks of transverse phonon modes along the [110] direction (dots). Solid lines denote the calculated phonon dispersion lines irrespective of their intensities. Calculations and measurements were performed for/in the Brillouin zones adjacent to *(c)* $\tau = (2, 2, 0)$, *(d)* $\tau = (3, 1, 0)$ and *(e)* $\tau = (0, 0, 3)$.

experimental results with DFPT we observe also some systematic deviations: for instance, the observed *a-b*-plane polarized transverse dispersion at about 10 meV [Fig. 1(d)] deviates from the prediction for $h \geq 0.3$ [in $q = (h, h, 0)$] in that the first transverse optic (TO) mode exhibits a pronounced dip in the wave vector region where it interacts with the upward dispersing TA branch in the form of an anti-crossing. Note that branches of the same symmetry are not allowed to cross. However, they can have an exchange of eigenvectors, where the character of the upward dispersing mode is transferred to the branch at higher energies and vice versa. Apparently, DFPT does not perfectly describe this scenario for the TA branch. On the other hand, the exchange of eigenvectors necessitates the analysis of several phonon branches in order to follow a particular phonon character, which is here the one of the TA phonon.

For a more detailed comparison, we separate the the measured phonon energies according to the Brillouin zones in which the measurements have been performed. DFPT predicts that the TA *a-b*-plane polarized branch along the [110] direction has the best structure factor in the Brillouin zone adjacent to the reciprocal lattice vector $\tau = (2, 2, 0)$. Regarding the first TO mode we get best scattering intensities around $\tau = (3, 1, 0)$. Raw data are shown in Figs. 2(a) and (b) and compared qualitatively to Gaussians representing the calculated structure factors. An summary of observed phonon peaks and predicted intensities is given in panels (c)-(e). We see that the observed energies nicely follow the TA dispersion up to the wave vector of the anti-crossing with the TO branch [Fig. 2(c)]. Near the zone boundary, the low-energy intensity also agrees well with experiment [Fig. 2(a)]. A second phonon peak, however, is observed at 10 meV instead of the predicted 12 meV. Measurements in the Brillouin zone around the reciprocal lattice vector $\tau = (3, 1, 0)$ corroborate this result [Fig. 2(b)]: The energy of the first TO mode agrees well with the prediction at the zone center but the scattering intensities deviate from the calculated ones at the zone boundary and intermediate wave vectors [Figs. 2(b),(d)]. We note that within our experimental resolution we could not see any energy broadening of the observed phonon peaks. As the character of the TA branch apparently does not disperse as high as predicted at the zone boundary, we believe that deviations from theory are related to a slightly overestimated strength of the anti-crossing rather than effects related to ASOC.

One short-coming of the prediction based on the kinetic theory of CePt$_3$Si [16] is that the detailed phonon polarization pattern is not taken into account. Hence, we investigated also the *c*-axis polarized phonon modes dispersing along the [110] direction to look for possible anomalies. Fortunately, we found a Brillouin zone, in which up to three different modes including the TA one have detectable intensities and can still be assigned unambiguously to the dispersion lines. Raw data were already shown in Fig. 1(a) and a more comprehensive comparison with DFPT for $Q = (h, h, 3)$ is given in Fig. 2(e). Note that DFPT predicts a rather complicated dispersion. In particular, the first TO branch shows near the zone center a downward dispersion and, after anti-crossings with the acoustic modes, disperses upwards towards the zone boundary as the nominal lowest energy TA mode. However, our experimental results imply a different scenario: here, the dispersion of the first TO mode is rather flat and for $h \geq 0.3$ it merges with the TA mode. Similar to the situation for the *a-b*-plane polarized transverse modes, it seems that DFPT overestimates the interaction between the various phonon branches.

It is well-known that the treatment of *4f* electrons in density-functional-theory is delicate because of their mixed character between being localized and itinerant. In order to judge the relevance of *4f* electrons for this particular case and for the dispersion in general, we performed another

DFPT calculation now for LaPt$_3$Si, where only *d* electrons are involved. In order to achieve the same electronic configuration as in CePt$_3$Si we added one additional *d* electron to LaPt$_3$Si. Furthermore, we used the experimental lattice constants and internal parameters for a realistic treatment of CePt$_3$Si (see the Appendix for more details).[2] Finally, phonon energies were obtained replacing the mass of the La ion with that of Ce. Nevertheless, we call it the "La calculation" for simplicity. Comparison between the two calculations for branches along the [110] direction are shown in Fig. 3. From the small differences in the lattice constants (see table 1) we expect harder energies for the *a-b*-plane polarized phonons modes and softer energies for *c*-axis polarized modes in the Ce-calculation. Many of the differences in Fig. 3 are indeed explained by this. Regarding our experimental results discussed above, the La calculation seems to do a slightly better job: the anti-crossing for the *a-b*-plane polarized branches [see experimental results in Fig. 2(a), (b)] is less strong and involves only the lowest two branches [Fig. 3(a)]. For c-axis polarized modes, it predicts a flat dispersion [Fig. 3(b)] of the first TO mode in agreement with our observations. We note here that we also performed DFPT calculations for CePt$_3$Si including spin-orbit coupling (SOC) (no results shown). The lattice constants are somewhat in between the optimized ones without SOC and the experimental ones but closer to the former (see Table 1). However, there is no particular change in the phonon dispersion apart from the slight softening because of the change in lattice constants.

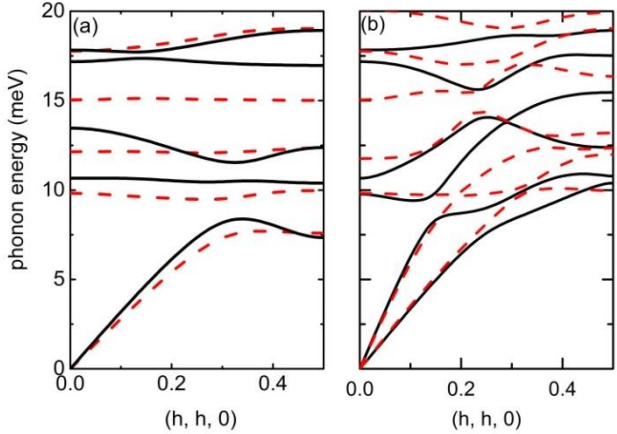

**Figure 3:** (color online) Comparison of calculated dispersions along the [110] direction for modes with *(a) a-b-*plane polarization and *(b) c-*axis polarization for CePt$_3$Si (solid lines) and LaPt$_3$Si (dashed lines). All wave vectors are given in reciprocal lattice units.

To summarize, we observed deviations in the phonon dispersion with regard to ab-initio predictions. However, we find the best agreement with experiment for the La calculation without *f* electrons and conclude that the differences do not originate in any sort of SOC. More likely, our calculations overestimate the exchange between different phonon branches, particularly in the presence of *4f* electron states.

## IV. Discussion and Conclusions

We provide the first detailed measurement of the lattice dynamical properties of CePt$_3$Si and make a detailed comparison with *ab-initio* calculations. We focused on TA phonon modes along the [110] direction motivated by the prediction of Kohn anomalies because of ASOC in the non-centrosymmetric crystal structure. The clearest deviation from lattice dynamical calculations based on DFPT was found for transverse *a-b*-plane polarized modes along the [110] direction. Here, the first TO mode predicted to have a flat dispersion features a pronounced dip. However, we found an improved agreement with a modified DFPT model neglecting the particular character of the one Ce *4f* electron. Hence, the observed deviations seem to result from an overestimated interaction strength between phonon branches of the same symmetry, which is amplified taking *4f* electrons into account.

We note that the predictions for Kohn anomalies for the ab-plane TA mode [16] were based on the assumption of constant electron-phonon coupling matrix elements and disregarding the exact phonon displacement patterns. On the other hand, investigations in other materials showed that the coupling between phonons and the electronic subsystem can depend on the type of atomic displacements corresponding to the electronic states involved at the Fermi surface[17]. Thus, it is possible that effects related to ASOC appear in different branches. We already accounted for this possibility by investigating also the lowest-energy *c*-axis polarized branches along the [110] direction but did not find any evidence for ASOC induced anomalies.

To summarize, we have evidence for neither Kohn anomalies nor increased line widths in any of our measurements. The fact that we did not find anomalies due to ASOC indicates that they are too small to be observed and, further, imply that the ASOC plays less a role in NCS superconductors than commonly believed. This is surprising, since on general grounds we expect to observe a strong and characteristic feature of a Rashba-type spin-orbit interaction. In most of the cases when $\vec{d} \parallel \vec{g}$ ($\vec{d}$: triplet part of the pair potential; $\vec{g}$: coupling of the ASOC [15]), the strength $\alpha$ of the ASOC is proportional to the strength of the triplet contribution of the superconducting order parameter. Even though predictions exist [15], the singlet-to-triplet ratio in CePt$_3$Si is unknown. Since we do not observe any clear fingerprint of the ASOC in the phonons, the triplet contribution seems to be small, maybe less than 10%. This would imply that the superconducting gap function reveals only point nodes rather than line nodes on one Fermi surface, while the other one is fully gapped [15]. Furthermore, all gauge modes and the Leggett mode, resulting from the Josephson effect between the two Fermi sheets, will be less unconventional for a small triplet contribution [13].

### Acknowledgements
S.K. and F.W. were supported by the Helmholtz young investigator group under contract VH-NG-840.

# Appendix

Here, we give more details about the performed ab-initio calculations and include results on the optimized structural parameters as well as the calculated electronic band structure.

## Technical details of the ab-initio theory

The calculations were performed within density-functional theory and the mixed-basis pseudopotential method,[18] which employs a combination of local functions and plane waves for the representation of valence states. The local density approximation for the exchange-correlation energy was used.[19] Phonon properties were calculated using the linear response technique[20] as implemented in the mixed-basis pseudopotential method.[21] Norm-conserving pseudopotentials were constructed from all-electron relativistic atom calculations according to the scheme of Vanderbilt.[22] $5s$ and $5p$ semi core states of La, Ce, and Pt were treated as valence states. The La pseudopotential was constructed without an explicit $f$ component by using an atomic configuration with empty $4f$ orbitals. All pseudopotentials included non-linear core corrections.

For CePt$_3$Si, we performed calculations both without and with spin-orbit coupling (SOC). SOC can be easily incorporated within the pseudopotential approach [23] and was treated fully self-consistently in the ground-state calculations.[24] For Brillouin zone summations a tetragonal $16 \times 16 \times 12$ **k**-point mesh was used in combination with the standard smearing technique employing a Gaussian broadening of $0.2$ eV. Phonon properties were calculated with the linear-response technique on a $4 \times 4 \times 2$ mesh. Dynamical matrices at arbitrary wave vectors were then obtained by a standard Fourier interpolation technique.

## Structural properties and electronic structure of CePt$_3$Si in density-functional theory

CePt$_3$Si crystallizes in a tetragonal crystal structure lacking inversion symmetry (space group P4mm) because of a missing mirror plane (0 0 1/2). Within this work, two different sets of calculations were used both performed within DFT. The first uses the electronic structure of LaPt$_3$Si and the experimental structural data published earlier.[2] The $f$-electron of the Ce-atoms was included by "doping" the system with an additional electron and compensating it by a homogeneous background charge. Phonon energies were corrected by using the mass of the Ce atom instead of the La one.

The second calculation used in this work uses the electronic structure of CePt$_3$Si, i.e. the $f$-electron was considered explicitly. The structure has been optimized within the local density approximation. Compared to the experimental structure,[2] the tetragonal axes are underestimated by approximately 2.2% while the c-axis is around 0.6% larger in DFT than the experimental value (see Table 1). The calculated unit cell volume is therefore 3.8% smaller than the experimental one. The internal parameters obtained in this optimized calculation are in good agreement with the experimental values. We also performed calculations

| Unit cell | a (Å) | c (Å) | a/c | V (Å³) |
|---|---|---|---|---|
| **OPT** (CePt$_3$Si) | 3.983 | 5.475 | 1.376 | 86.857 |
| **SOC** (CePt$_3$Si) | 3.990 | 5.465 | 1.369 | 87.003 |
| **EXP** (LaPt$_3$Si) | 4.072 | 5.442 | 1.337 | 90.235 |
| z-value of site | La/Ce | Pt(1) | Pt(2) | Si |
| **OPT** (CePt$_3$Si) | 0.1398 | 0.6534 | 0 | 0.4177 |
| **SOC** (CePt$_3$Si) | 0.1425 | 0.6542 | 0 | 0.4180 |
| **EXP** (LaPt$_3$Si) | 0.1468 | 0.6504 | 0 | 0.4118 |

**Table 1:** Structural parameters used in the ab-initio calculations: "OPT" denotes the structure optimized within LDA and used for explicitly including Ce atoms. In a second step, we also included spin-orbit coupling for which results are labeled as "SOC". Experimental lattice constants and internal parameters as reported in Ref. 4 were used for a reference calculation of LaPt$_3$Si labeled "EXP".

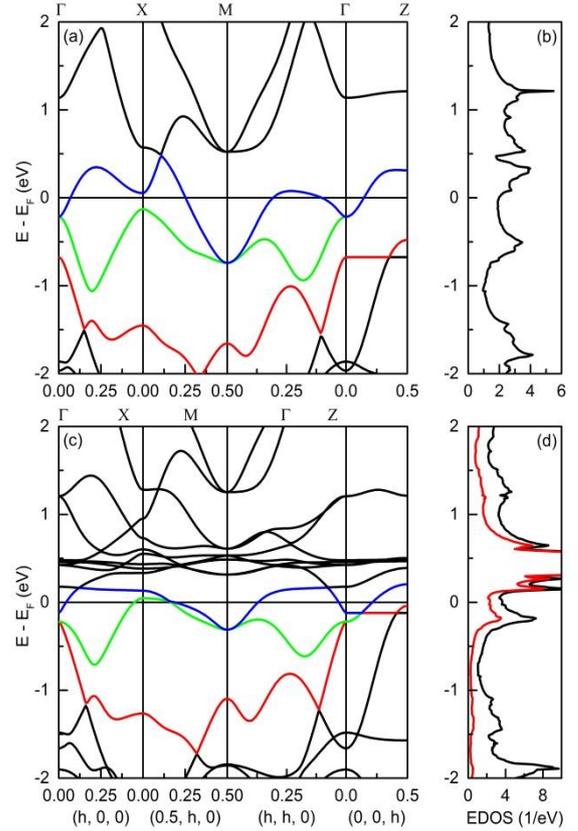

**Figure 4:** (color online) Calculated electronic band structure for *(a)* LaPt$_3$Si with one additional electron and for *(c)* CePt$_3$Si considering the 4f-electron of Ce explicitly (without SOC). Electronic bands contributing to the Fermi surface are shown in red (α-band), green (β-band) and blue (γ-band). Figures on the right *(b),(d)* show the corresponding electronic densities of states. The Ce's 4f-electron partial eDOS is shown in red in panel *(d)*. Wave vectors are given in reciprocal lattice units

including spin-orbit coupling (SOC). Here, the structural parameters are close to those without SOC and the changes of about 0.2% bring it only slightly closer to the experimental structure.

For comparison we show results for the electronic structure along several high symmetry directions for both La- and Ce-calculations in Fig. 4. We label the three bands crossing the Fermi energy $E_F$ in the latter with α (red), β (green) and γ (blue) as done in Ref. [3]. Going from the La- to the Ce-calculation we see a clear shift of the α band towards $E_F$, which actually does not cross $E_F$ in the former. The other two bands are also shifted upwards (β band) and compressed (β and γ bands) in energy resulting in a clear change of the Fermi surface. Very prominent is a peak in the electronic density of states (EDOS) in the Ce-calculation centred on 0.4 eV above $E_F$. According to the partial EDOS, the peak is related to unoccupied *4f* states of the Ce with flat dispersions and the states at $E_F$ are also of predominantly *4f* character. Generally, results of the Ce calculation are in good agreement with the ones presented in Ref. [3].